\documentclass{iau}
\usepackage{graphicx}

\title[The Keplerian orbit of G2] 
{The Keplerian orbit of G2}

\author[Meyer et al.]   
{L. Meyer$^1$, A. M. Ghez$^1$, G. Witzel$^1$, T. Do$^2$, K. Phifer$^1$,\\ B. N. Sitarski$^1$, M. R. Morris$^1$, A. Boehle$^1$, S. Yelda$^1$,\\ J. R. Lu$^3$, E. Becklin$^1$
}

\affiliation{$^1$ Department of Physics and Astronomy, University of California,\\
    Los Angeles, CA 90095-1547, USA,  email: {\tt leo@astro.ucla.edu} \\[\affilskip]
$^2$Dunlap Institute for Astronomy and Astrophysics, University of Toronto,\\
Toronto M5S 3H4, Ontario, Canada,\\
$^3$ University of Hawaii, Institute for Astronomy,  Honolulu, HI 96822, USA
}

\pubyear{2013}
\volume{303}  
\pagerange{119--126}
\jname{The Galactic Center: Feeding and Feedback in a Normal Galactic Nucleus}
\editors{A.C. Editor, B.D. Editor \& C.E. Editor, eds.}
\begin{document}

\maketitle

\begin{abstract}
We give an update of the observations and analysis of G2 -- the gaseous red 
emission-line object that is on a very eccentric 
orbit around the Galaxy's central black hole and predicted to come within 2400 $R_S$ in early 2014. During 2013, the laser guide star adaptive optics systems on the W. M. Keck I and II 
telescopes were used to obtain three epochs of spectroscopy and imaging at the highest spatial 
resolution currently possible in the near-IR. The updated orbital solution derived from radial velocities in addition to 
Br-$\gamma$ line astrometry is consistent with our earlier estimates. Strikingly, even $\sim 6$ months before pericenter passage there is no perceptible deviation from a Keplerian orbit. We furthermore show that a proposed ``tail" of G2 is likely not associated with it but is rather an independent gas structure. We also show that G2 does not seem to be unique, since several red emission-line objects can be found in the central arcsecond.   Taken together, it seems more likely that G2 is 
ultimately stellar in nature, although there is clearly gas associated with it. 
\end{abstract}

\section{Introduction}

Recently, Gillessen et al. (2012,  2013a,b) reported the discovery of a 3 M$_{\rm Earth}$ gas cloud plunging toward the supermassive black hole (SMBH) at the Galactic Center (GC) with a predicted closest approach of only 2400 times the radius of the event horizon. If this object -- called G2 -- is indeed a gas cloud, it would be ripped apart by the tidal forces of the SMBH during closest approach and then accreted (Burkert et al. 2012; Schartmann et al. 2012; Anninos et al. 2012; Shcherbakov 2013; Abarca et al. 2013). While the identification of the source as a gas cloud is controversial and many alternative models containing a central stellar source have been proposed (e.g., Murray-Clay \& Loeb 2012; Miralda-Escude 2012; Eckart et al. 2013; Scoville \& Burkert 2013; Ballone et al. 2013), all the observations indicate that low density gas associated with this object is being tidally disrupted (Gillessen et al. 2012, 2013a,b; Phifer et al. 2013). In the case of a pure gas cloud, models predict radio shocks to begin 7 to 9 months ahead of periapse passage of the center of mass (Narayan et al. 2012; Sadowski et al. 2013). No such shocks have been observed to date. Regardless of the underlying nature of G2, most models predict that the disrupted material should eventually be accreted onto the SMBH, possibly as an observable event (e.g., Fragile et al. 2013). This could provide a rare opportunity to follow a predicted accretion event, potentially teaching us about black hole accretion physics.

\section{Kinematics of G2 and its ``tail" emission}

The first orbit for G2 was derived from data based on L'-band (3.8 $\mu$m) astrometry and Br-$\gamma$ radial velocities (Gillessen et al. 2012). It was then shown by Phifer et al. (2013) that a more accurate solution is found by using the Br-$\gamma$ emission for astrometry as well, shifting the predicted time of closest approach from mid-2013 to early 2014 (see first column of table~\ref{tab1}). Here, we report an update to Phifer et al.'s (2013) solution by incorporating new 2013 data. 

The new data set consists of three epochs with 36 frames taken on May 11--13, 28 frames taken on July 25--27, and 32 frames taken on August 11--13 with the AO-fed Integral Field Spectrograph OSIRIS. Each frame has an integration time of 15 mins. In July and August the narrow-band Kn3 filter, which is centered on the Br-$\gamma$ hydrogen line (2.1661 $\mu$m) was used, while in May the Kbb broadband filter was selected. Both set-ups led to a spectral resolution of R $\sim$ 3600. The data reduction, extraction of radial velocities, and absolute astrometric positions were carried out in the same way as in Phifer et al. (2013).

\begin{figure}[tb]
\begin{center}
 \includegraphics[width=7.cm]{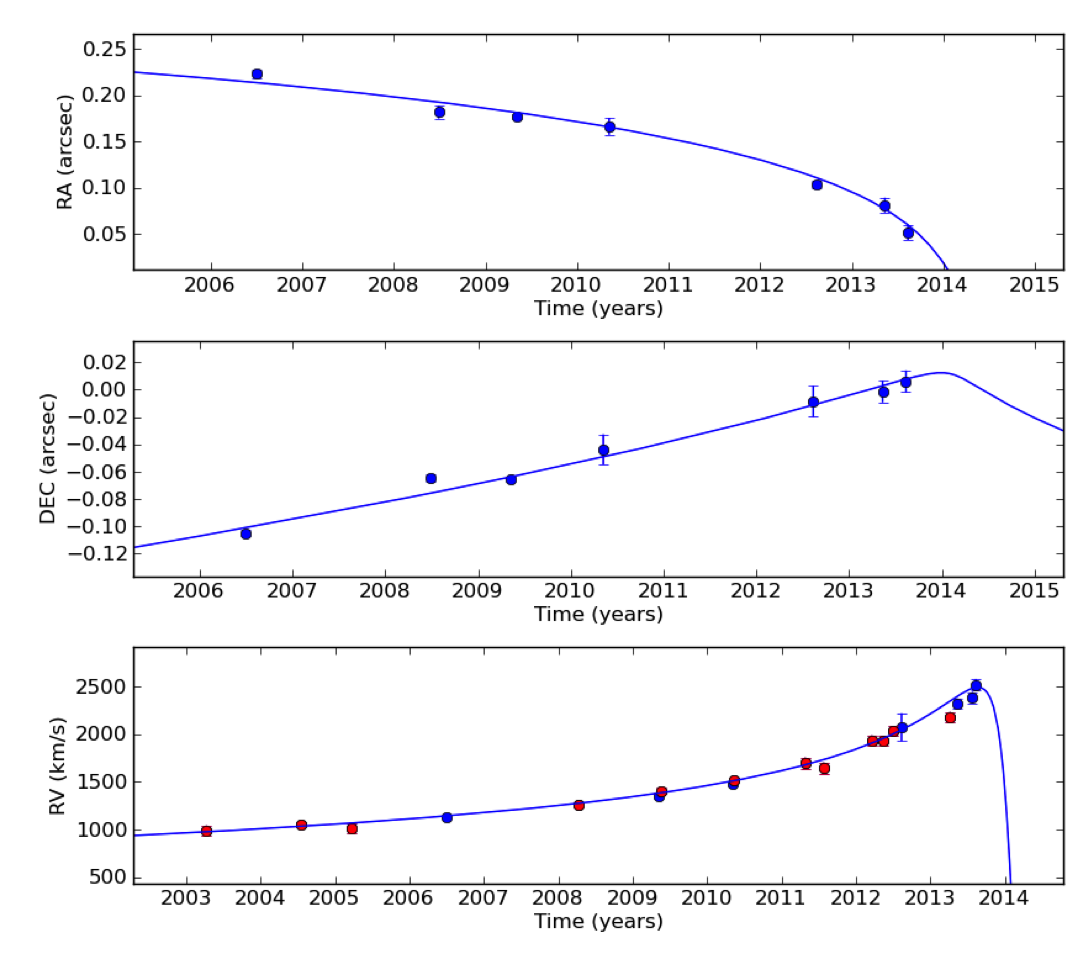}
 \includegraphics[width=6.cm]{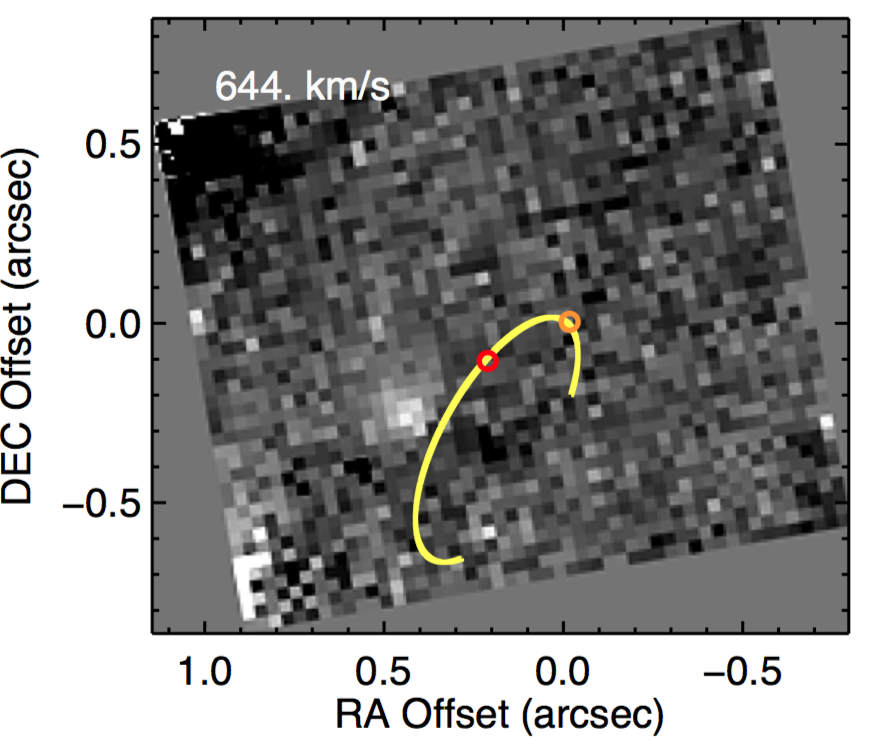} 
 \caption{The Keplerian orbit of G2. \textit{Left:} G2's astrometric positions and RVs as a function of time, with the best-fit Keplerian orbit superposed as a solid line. The blue points are from our OSIRIS data, the red RV points are from Gillessen et al. (2013b). Even 6 months before pericenter passage, a purely Keplerian orbit is sufficient to describe the motion of G2. \textit{Right:} A slice of OSIRIS' data cube corresponding to a velocity of 644 km/s. Further downstream in G2's orbit (G2 itself is not visible since its RV is different, its position in 2006 is marked by the red, lower left circle; Sgr~A*'s position is the orange, upper right circle) a gas structure becomes apparent that has been interpreted as tail emission based on a position-velocity diagram. However, spatially G2's orbit (solid line) does not seem to be associated with this structure.}
   \label{fig1}
\end{center}
\end{figure}

The third column in table~\ref{tab1} shows our current best-fit orbital solution. Both near-IR groups have converged to consistent orbits. Figure~\ref{fig1} shows the astrometric\footnote{The astrometric position is defined as the flux-weighted centroid.} and RV data points together with the best-fit. It is interesting to note that a purely Keplerian orbit is sufficient to describe the motion of G2 . Given the possible interpretation of G2 as a pure gas cloud, it is unclear how a perfectly Keplerian orbit can be maintained up to 6 months before pericenter. A central (stellar) core within the cloud seems a straightforward explanation of the observed dynamics. 

The right panel of Figure~\ref{fig1} shows the Br-$\gamma$ emission at a lower RV of $\sim 650$ km/s which was claimed to fall on G2's orbit further downstream and was therefore interpreted as G2's tail emission (Gillessen et al. 2012). While the RV does coincide with G2's orbit, this gas feature does not fall on G2's orbit spatially (see also Phifer et al. 2013). In a position-velocity diagram (Gillessen et al. 2012, 2013a,b) this fact gets masked by the finite width of the curved slit used to extract RVs along the orbit.

\begin{table}
  \begin{center}
 \caption{Orbital solutions$^1$ for G2 based on Br-$\gamma$ astrometry and radial velocities}
\label{tab1}
 \begin{tabular}{lccc}\hline 
 Param & { UCLA 2012$^{2}$} & { MPE 2013$^{3}$} & {\bf UCLA 2013} \\ \hline
ecc  & $0.981 \pm 0.006$ & $0.970 \pm 0.003$ & $0.965 \pm 0.011$  \\ 
Incl [deg]  & $121 \pm 3$ & $118 \pm 2$ & $113 \pm 3$  \\ 
$\Omega$ [deg]  & $56 \pm 11$ & $82 \pm 4$ & $77 \pm 10 $  \\ 
$\omega$ [deg] & $88 \pm 6$ & $97 \pm 2$ & $92 \pm 4$  \\ 
T$_0$ [year]  & $2014.21 \pm 0.14$ & $2014.25 \pm 0.06$ & $2014.21 \pm 0.13$  \\ 
P [years]  & $276 \pm 111$ & $391 \pm 66$ & $264 \pm 139$  \\ \hline
  \end{tabular}
 \end{center}
 \scriptsize{
 {\it Notes:}\\
  $^1$The mass of the BH is $3.8\times 10^6 M_\odot$ and the distance is $7.6$ kpc. \\
  $^2$From Phifer et al. (2013).\\
  $^3$From Gillessen et al. (2013b).}
\end{table}

\section{G2 does not seem to be unique}

\begin{figure}[htb]
\begin{center}
 \includegraphics[width=7.75cm]{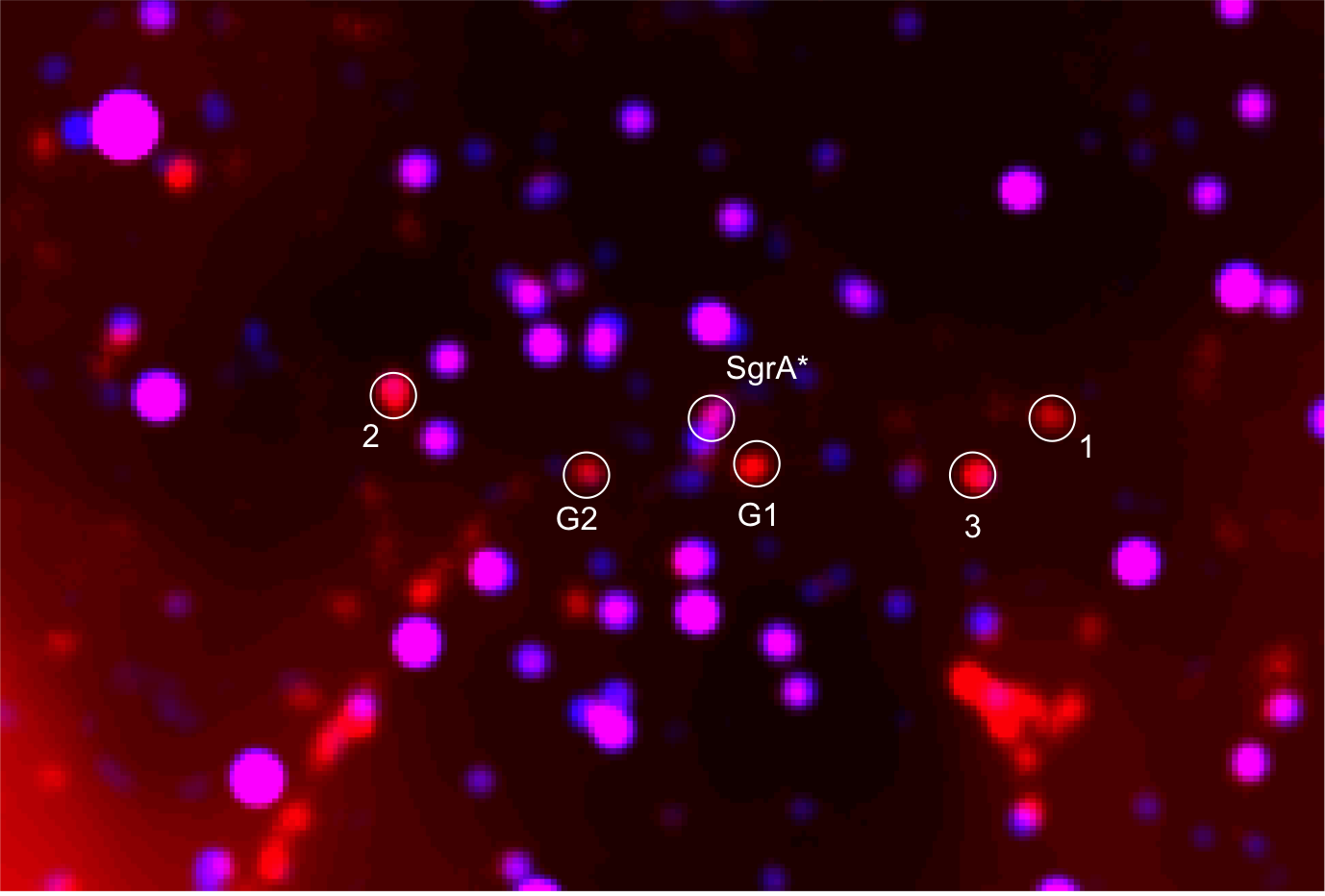}
 \includegraphics[width=5.25cm]{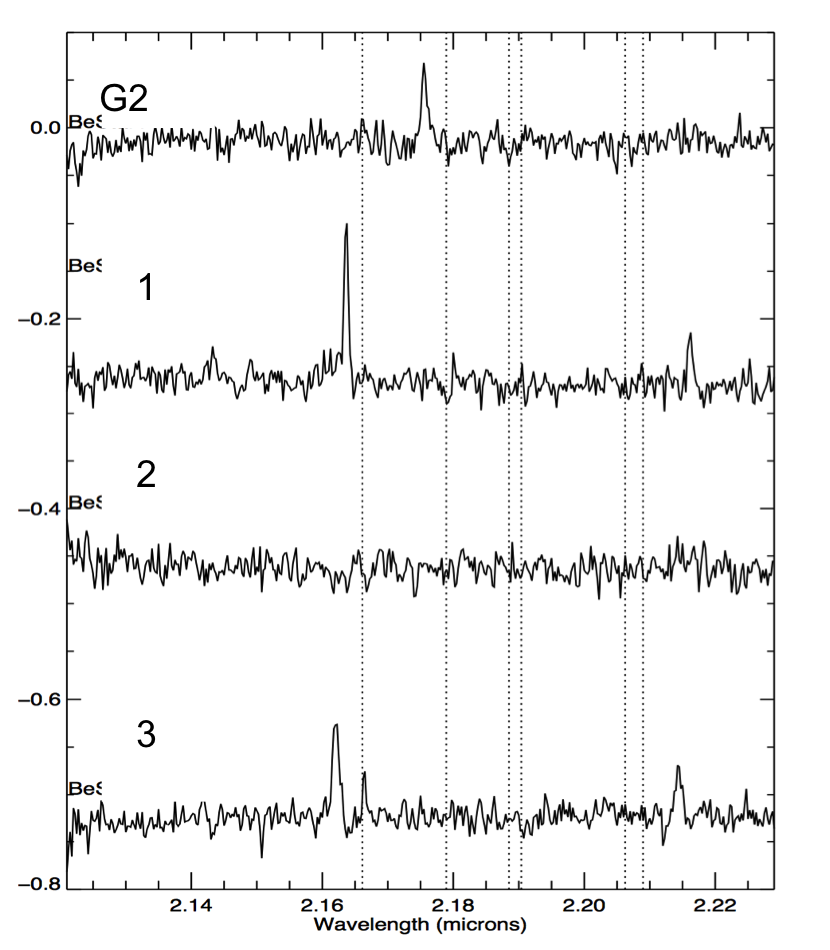} 
 \caption{G2 in context: red emission-line sources at the GC. \textit{Left:} A K- and L-band composite (2.1 $\mu$m and 3.8 $\mu$m)  of the central 2.4" x 1.6" of the Galaxy. Obvious infrared excess sources are marked with the circles along with G2 and Sgr~A*. \textit{Right:} Corresponding spectra. The vertical dashed lines mark the rest wavelengths of typical stellar lines. Sources 1 and 3 show hydrogen emission with no other features, very similar to G2. }
   \label{fig2}
\end{center}
\end{figure}

An important context in the discussion of G2's properties and nature is the abundance of similar sources in the central arc-secs of our Galaxy. Is G2 the only red emission-line object at the very center? Figure~\ref{fig2} (left panel) shows a K- and L-band image overlay (2.1 $\mu$m and 3.8 $\mu$m). Quite a few very red sources similar to G2 clearly stand out (see also Eckart et al. 2013). The right panel plots spectra for some of these sources. A subset shows significant Br-$\gamma$ emission and no absorption features typical for late-type giants. While G2, in contrast to the other sources, shows substantial RV changes and an evolution of the emission-line's FWHM, it seems to be one of many similar sources. It is mainly noticeable because its pericenter passage is happening in 2014.

\section{No G2 K-band flux above the detection limit}

\begin{figure}
\begin{minipage}[c]{6.5cm}
 \includegraphics[width=6.cm]{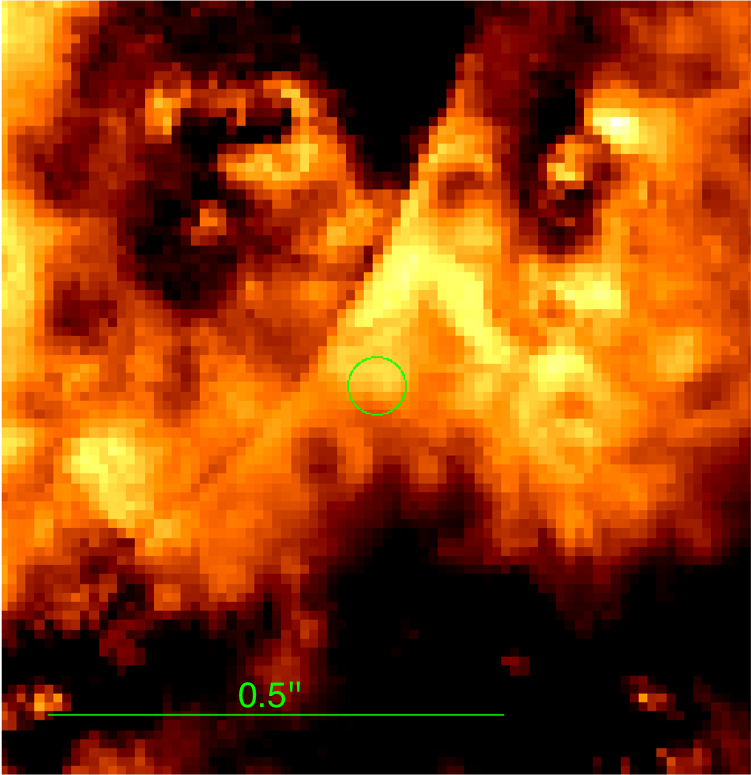}
\end{minipage}
\begin{minipage}[c]{6cm}
 \caption{No significant K-band detection of G2. Average detection residuals from 2008 to 2012 are shown. The residuals have been calibrated photometrically and centered on G2 in each epoch. An existing residual point source not detected in a single epoch would add up at the position marked by the green circle. This is not the case. The flux inside the green circle does not show any significant excess with respect to the halo noise of the bright IRS16 sources. The visible curve stretching from the lower left to the upper right marks the edge of the subtracted PSF from IRS16C. Clearly, limited PSF knowledge is the dominating noise source.}
 \label{fig3}
 \end{minipage}
\end{figure}
 
Both Gillessen et al. (2012) and Phifer et al. (2013) reported no detection of G2 in the K-band, with the latter paper stating a limit of K$_{\rm mag} = 20$ for a point source. Eckart et al. (2013; this volume) on the other hand find a G2 K-band counterpart with a K$_{\rm mag} = 19$ in both VLT and Keck data. Here, we do not confirm this detection in the Keck data. While there is some residual extended flux at the position of G2 (as inferred from the Br-$\gamma$ based orbit), this flux level is not greater than typical contributions from the halos of the bright IRS 16 sources (see Fig.~\ref{fig3}). A residual map that combines many epochs from 2008 -- 2012, using the known motion of G2 to transform them into a G2 rest-frame, does not show a source at G2's position. The flux level inside a circular aperture is not above the level of test apertures placed at random positions. While it cannot be ruled out that some of the flux is indeed associated with G2, we don't see any evidence for this connection when halo noise clearly dominates the region in these low flux levels.

\textit{Acknowledgements:} Support for this work was provided by NSF grant AST 0909218 and the Levine-Leichtman Family Foundation.


\end{document}